# Few-picosecond pulse generation featuring ultrafast spectral dynamics in gain-switched surface-grating DFB lasers via impulsive optical pumping

Yihan Qi,[1,*] Fuyi Cao,[1] Hidekazu Nakamae,[1] Changsu Kim,[1,3] Masataka Kobayashi,[1] Cong Wang,[1] To-Fan Pan,[1] Shaoqiang Chen,[1,2] Takashi Ito,[3] AND Hidefumi Akiyama[1,3]

[1]*Institute for Solid State Physics, The University of Tokyo, 5-1-5 Kashiwanoha, Kashiwa 277-8581, Japan*
[2]*State Key Laboratory of Precision Spectroscopy, Department of Electronic Engineering, East China Normal University, Shanghai 200241, China*
[3]*LDseed Co., Ltd., 684 Fujisawa, Fujisawa 251-0052, Japan*
*Corresponding author: qi-yihan129@g.ecc.u-tokyo.ac.jp

**Abstract:** To investigate the physics of picosecond gain-switching dynamics in single-mode lasers under femtosecond optical pumping at room temperature, we designed and fabricated first-order surface-grating GaAs distributed-feedback (DFB) lasers with five systematically varied grating periods (120–124 nm), corresponding to lasing wavelengths of 825.7–849.5 nm (1.502–1.459 eV). The 124-nm-period device, closest to the quantum-well gain peak among the investigated devices, exhibited the highest output power and spectral bandwidth. Among all devices, the 122-nm-period DFB laser (838.2 nm, 1.480 eV) generated the shortest pulses, despite lasing at a higher photon energy and lower output power than the device closest to the gain peak. All devices exhibited characteristic down-chirp behavior that increased with excitation power. The shortest pulses had a chirped pulse width of 6.6 ps and a chirp rate of 0.13 meV/ps, whereas spectrally resolved measurements revealed a minimum pulse width of 3.8 ps (2.3 ps after deconvolution of the detection time resolution) near the central photon energy of the pulse spectrum. Numerical simulations revealed temporally and spatially resolved dynamics of photons, carriers, gain, and refractive index, reproducing the experimental results qualitatively and quantitatively. Furthermore, a mechanism for generating the shortest pulses at photon energies above the gain peak was identified and attributed to higher differential gain, saturation gain, and a higher transparency carrier density in the high-energy region of the gain spectrum. These experimental and theoretical results elucidate the intrinsic dynamics of picosecond pulse generation in gain-switched DFB lasers and provide design guidance for short-pulse generation and computational tools applicable to both optical and electrical pumping.

## INTRODUCTION

Gain-switched semiconductor lasers [1–5] are promising picosecond light sources that combine high efficiency, compactness, robustness, and easy integration with electronic and photonic systems. They are widely used in applications such as laser processing [6–8], multiphoton microscopy [9,10], medical laser surgery [11], and integrated photonic systems [12,13]. However, their internal nonlinear optical physics is often unpredictable because of complex nonlinear gain dynamics driven by nonequilibrium, high-density interactions among electron–hole carriers in semiconductor lasers [14–17].

Theoretical models, methods, tools, and computational capabilities have been significantly developed to deal with such problems [18–23]. However, theoretical calculations alone remain insufficient to predict the short-pulse limit and its necessary conditions in gain-switched semiconductor lasers. They still require parameter and model tuning using unambiguous experimental data. For this purpose, gain-switching experiments based on femtosecond impulsive optical pumping are best suited, as optical pumping is free from the modulation-bandwidth limitation of electrical pumping and enables delta-function-like impulsive excitation. It provides well-defined initial conditions and simplifies theoretical explanation. Moreover, impulsive pumping provides a direct pathway for analyzing gain-switching physics in the extreme nonlinear regime, both experimentally and theoretically. So far, gain-switching experiments with femtosecond impulsive optical pumping have been performed for Fabry–Perot (FP) lasers [24,25], vertical-cavity surface-emitting lasers (VCSELs) [26–28], and other types of semiconductor lasers [29,30], but not yet in distributed-feedback (DFB) lasers because of fabrication-related challenges.

Gain-switched DFB lasers offer unique advantages, including stable single-mode operation at a desired wavelength, high stability, and broad applicability [31–33]. For instance, Riecke *et al*. [34] used electrically driven 1060-nm DFB lasers to generate 70-ps gain-switched pulses and investigated the competition between Bragg modes and FP modes at different temperatures. Chen *et al*. extracted nearly 5-ps optical pulses from a 1550-nm DFB laser using spectral filtering [35,36]. Kobayashi et al. used a 1270-nm DFB laser with a direct-modulation bandwidth of 30 GHz to generate 5.3-ps pulses after chirp compensation [37]. However, impulsive optical pumping experiments have not been performed yet for DFB lasers, possibly because of the difficulty in systematically preparing DFB-laser samples for such experiments. Thus, fundamental experimental studies and comparable theoretical analyses of gain-switched short-pulse generation in DFB lasers remain lacking.

In this work, we designed and fabricated first-order surface-grating (SG) DFB lasers [38–42] with systematically varied grating periods to control the lasing wavelength. Using these devices, we investigated gain-switched short-pulse generation induced by impulsive optical pumping at room temperature. Our experiments revealed that lasing on the higher-energy side of the gain peak leads to significantly shorter gain-switched pulses. We obtained a short, chirped pulse width of 6.6 ps and a chirp rate of 0.13 meV/ps in an SG-DFB laser at room temperature. The shortest spectrally resolved pulse width was 3.8 ps (2.3 ps after deconvolution of a 3.0-ps detection time resolution). We performed simulations incorporating local photon, carrier, gain, and refractive-index dynamics along the cavity. The simulations captured the key experimental trends and demonstrated quantitative agreement with the measured pulse characteristics. By combining experimental and theoretical results, we elucidate the ultrafast physical mechanisms underlying the generation of the shortest pulses at photon energies higher than the gain center, which we attribute to higher differential gain, enhanced saturation gain, and a higher transparency carrier density at higher photon energies.

## STRUCTURE AND FABRICATION

Figure 1 illustrates the schematic of our SG-DFB lasers for optical pumping. SG-DFB lasers are easier to fabricate than buried DFB lasers because they do not require a second epitaxial regrowth. In addition, SGs enable controlled, strong optical coupling and efficient modulation. The lasers' epitaxial layers were grown via molecular beam epitaxy on an n-GaAs (100) substrate. The active layers comprised ten periods of multiple quantum wells (MQWs), consisting of 12-nm GaAs wells and 10-nm $Al_{0.33}Ga_{0.67}As$ barriers, sandwiched between $Al_{0.33}Ga_{0.67}As$ cladding layers. SGs with a duty cycle of 50:50, periods $\Lambda =120$–124 nm, and a width of 3.3 μm were first patterned using electron-beam lithography (EBL) (ELS-BODEN 100) and subsequently transferred onto a silicon dioxide ($SiO_2$) mask via inductively coupled plasma reactive ion etching (ICP-RIE) using a low-damage Spica ICP-RIE system. The grating pattern was then etched into the cladding layer to a depth of ~220 nm using atomic layer etching (ALE PlasmaPro 100). A $\lambda/4$ phase shift was introduced at a position 100 μm from the output facet in a 250-μm-long cavity to enhance output power and ensure single-mode operation.

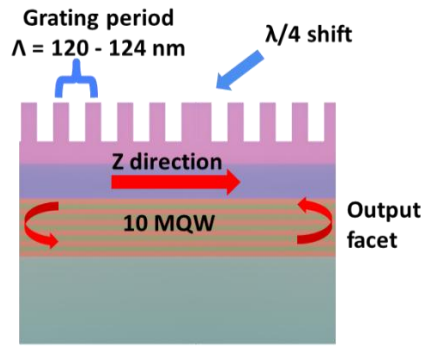

Fig. 1. Schematic cross-sectional view of the SG-DFB laser.

We then formed ridge waveguides along the *z*-axis with a width of 3.7 μm. The top surface was coated with an antireflective $SiO_2$ layer deposited via chemical vapor deposition to enhance optical pumping efficiency. The wafer was subsequently thinned and cleaved to form a cavity length of 250 μm, with uncoated as-cleaved facets that provide a reflectivity of 0.3. With this structure, the designed coupling coefficient ($\kappa$) was 4000 m$^{-1}$, corresponding to $\kappa L \approx 1$. We also fabricated ridge-waveguide FP lasers without the SG structure for reference.

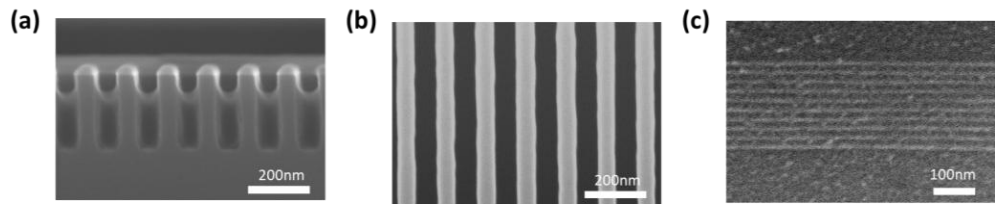

Fig. 2. SEM images of the (a) cross-section and (b) top view of the grating of the fabricated SG-DFB laser. (c) Active region comprising ten periods of GaAs quantum wells and $Al_{0.33}Ga_{0.67}As$ quantum barriers.

Figures 2(a) and 2(b) show scanning electron microscopy (SEM, Hitachi SU8230) images of the cross-section and top view of the grating formed on the ridge waveguide of an SG-DFB laser. Residues observed between the grating grooves correspond to the electron-beam resist that remained to protect the grating morphology during trench etching. This resist was subsequently removed after completing the ridge-waveguide fabrication. Figure 2(c) shows the SEM image of the active layer consisting of ten periods of 12-nm GaAs MQWs and 10-nm $Al_{0.33}Ga_{0.67}As$ barriers.

### EXPERIMENTAL SETUP

In this study, impulsive optical excitation was used to investigate the gain-switching pulse dynamics in single-mode SG-DFB semiconductor lasers, as shown in Fig. 3. A Ti:sapphire mode-locked laser (Tsunami) was used as the femtosecond optical excitation source, which had a pulse width of 200 fs, a central wavelength of 780 nm, a repetition rate of 80.8 MHz, and an average pumping power of up to 170 mW. A portion of the light was directed to a high-speed photodetector to serve as a trigger signal for a streak camera.

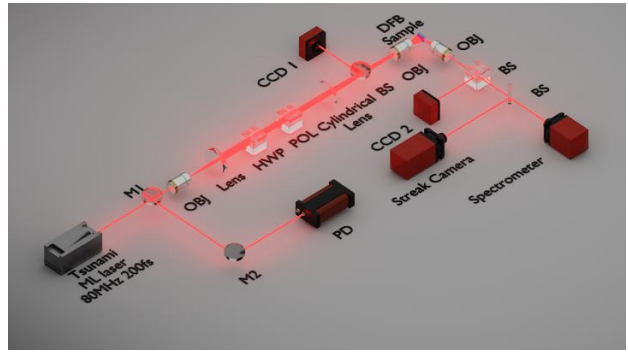

Fig. 3. Experimental setup for the optical-pumping measurement of the DFB laser. M: mirror; OBJ: objective; PD: photodetector; HWP: half-wave plate; POL: polarizer; CCD: charge-coupled device; BS: beam splitter.

An SG-DFB laser bar was mounted on a copper block at room temperature (300 K). Filament-shaped pump pulses were focused on the ridge waveguide of the SG-DFB laser. To achieve a filament-shaped optical pumping profile, a cylindrical lens was combined with an objective lens for beam shaping. The profile was optimized for uniform pumping along the ridge waveguide. The time-integrated optical spectra of output pulses from a facet of the SG-DFB laser were measured using a spectrometer (SpectraPro 750) with a resolution of 0.02 nm. Time-resolved spectra were recorded using a synchronously scanned streak camera (Hamamatsu C10910) equipped with a spectrometer, providing a time resolution of ~3.0 ps and a spectral resolution of 0.8 nm.

**EXPERIMENTAL RESULTS**

Figure 4(a) presents the time-integrated emission spectra of both DFB and FP lasers, measured using a spectrometer at an average optical pumping power of 170 mW. The FP lasing spectrum (red curve) had a broad spectral width of ~18.28 meV (10.78 nm), a longitudinal mode spacing of 0.71 meV (0.41 nm), and a peak wavelength of 1.460 eV (849 nm). In contrast, the DFB laser with a grating period of 124 nm (orange curve) operated in single mode, with a central wavelength of 1.459 eV (849.5 nm) and a much narrower spectral width of 1.51 meV (0.88 nm). This DFB laser also exhibited the highest peak intensity because its lasing wavelength was closest to the FP laser's peak wavelength, where the optical gain was maximized.

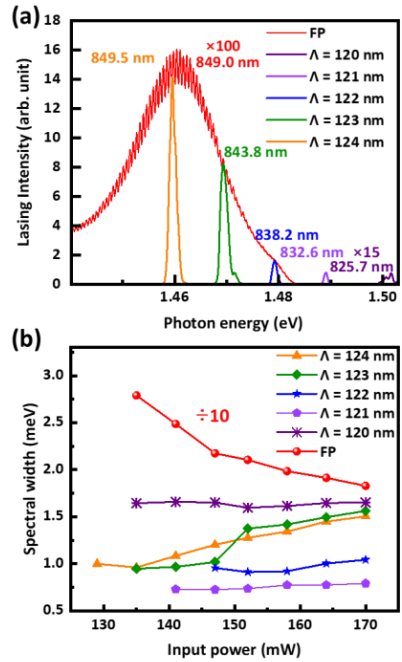

Fig. 4. Spectral measurements of DFB lasers with different grating periods ($\Lambda$) and an FP laser at a pumping power of 170 mW. (a) Emission spectra. (b) Spectral widths.

As the grating period decreased from 124 to 120 nm (dark purple line), the lasing wavelengths decreased from 849.5 to 825.7 nm (1.502–1.459 eV), and the peak intensity progressively decreased. This behavior could be attributed to the gradual detuning of the lasing wavelength from the FP laser's central wavelength, which reduced optical gain. The 120-nm-period DFB laser exhibited multiple side modes and a lower peak intensity due to large detuning from the gain center, thereby weakening the DFB feedback.

Figure 4(b) illustrates the spectral width as a function of pumping power. The FP laser exhibited a spectral width of 27.90 meV (16.56 nm) at 135 mW, which further narrowed to 18.28 meV (10.78 nm) at a higher pumping power of 170 mW because of the increased gain competition among longitudinal modes. The DFB lasers consistently demonstrated spectral widths narrower than those of the FP laser. Moreover, devices with smaller grating periods led to an even further narrowing of the spectral width, except for the 120-nm-period DFB laser (dark purple line), which showed a broad spectral width due to multimode lasing. Additionally, spectral broadening was observed in all DFB lasers at high pumping power.

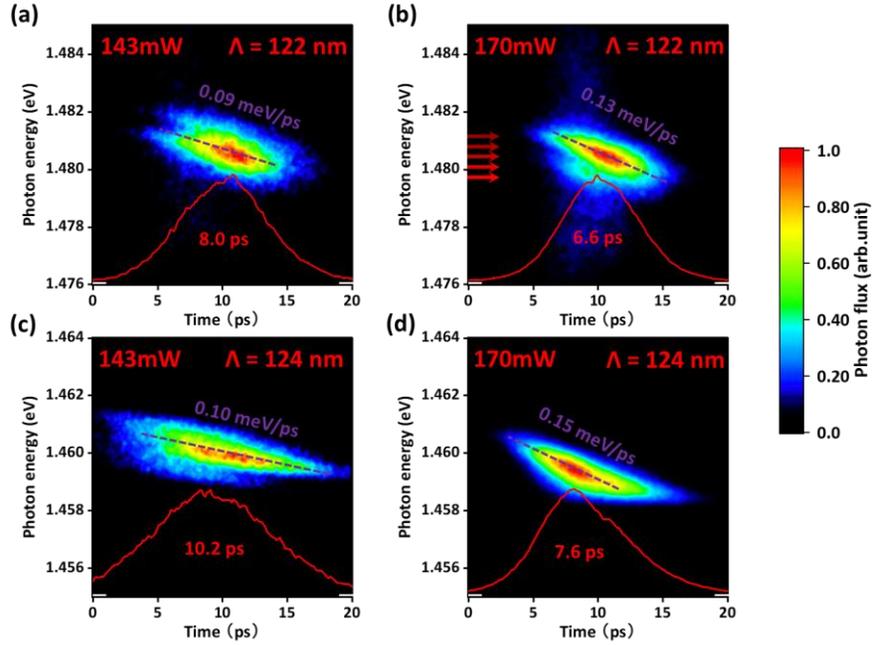

Fig. 5. Measurement of pulse dynamics for DFB lasers with a grating period of (a, b) 122 nm and (c, d) 124 nm. Spectral evolution of the gain-switched pulse under pumping powers of (a, c) 143 mW and (b, d) 170 mW measured using a streak camera. The color bar represents the normalized photon flux.

Figure 5 presents the time-resolved spectra of pulses from the 122-nm- and 124-nm-period DFB lasers, measured using a streak camera at different optical pumping powers. As shown in Fig. 5(a), the SG-DFB laser with a 122-nm grating period, pumped at a lower power of 143 mW, exhibited a pulse width (full width at half-maximum [FWHM]) of 8.0 ps and a down-chirp rate of 0.09 meV/ps (0.05 nm/ps), where the photon energy continuously decreased with time. As the excitation power increased to 170 mW, the pulse width decreased to 6.6 ps, whereas the chirp rate increased to 0.13 meV/ps (0.07 nm/ps). These overall spectral dynamics indicated that stronger excitation produces higher optical gain and faster carrier depletion, leading to enhanced chirp. In Figs. 5(c) and 5(d), the output pulses of the SG-DFB laser with a grating period of 124 nm also showed a similar behavior. However, at pumping powers of 143 and 170 mW, the pulse widths (10.2 and 7.6 ps, respectively) were broader than those of the 122-nm-period DFB laser, despite the higher output intensity of the 124-nm-period device. In addition, the corresponding chirp rates were ~0.10 meV/ps (0.06 nm/ps) and 0.15 meV/ps (0.09 nm/ps), respectively, which were higher than those of the 122-nm-period device.

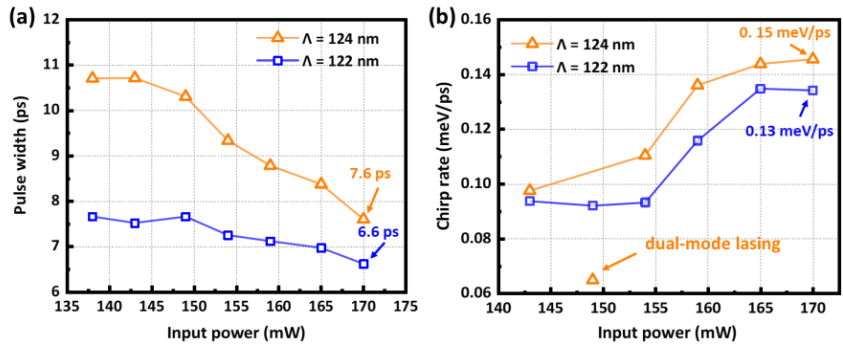

Fig. 6. (a) FWHM pulse width as a function of pumping power for DFB lasers with different grating periods (Λ) and for an FP laser. (b) Chirp rate extracted via linear regression from the time-resolved spectral measurements. The data point for Λ = 124 nm at 149 mW deviates because of dual-mode lasing and is omitted from the guide-to-the-eye curves.

Figure 6(a) illustrates the variation of pulse width with pumping power for DFB lasers with grating periods of 122 and 124 nm, measured using a streak camera. In both cases, the pulse width was clearly dependent on pumping power, with higher pumping power yielding shorter pulse widths. In the pumping power range of 138–170 mW, the 122-nm-period DFB laser exhibited a shorter pulse width than the 124-nm-period device.

Figure 6(b) illustrates the power dependence of the down-chirp rate, obtained via linear regression of the time-resolved spectra. The chirp rate increased monotonically with pumping power. An exception was the low chirp rate at 149 mW for the 124-nm-period device, which was due to dual-mode lasing and could thus be neglected. This overall trend should be attributed to shorter pulse durations and larger refractive-index changes at higher carrier densities. In addition, the 124-nm-period DFB laser consistently exhibited a higher chirp rate than the 122-nm-period device over the investigated power range, reflecting a stronger carrier-density change.

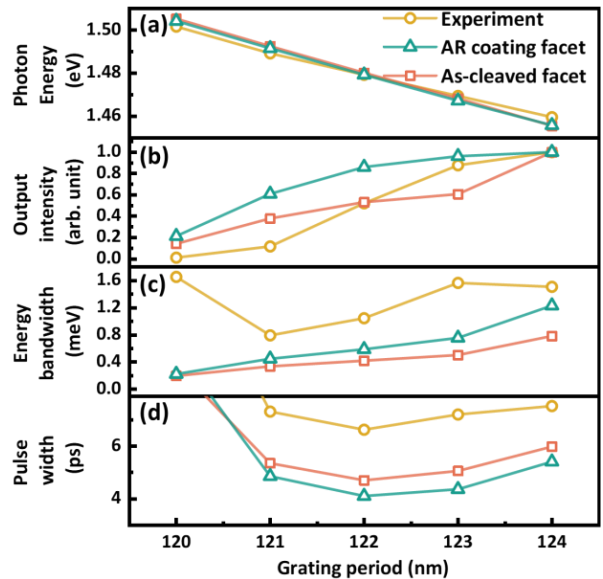

Fig. 7. (a) Lasing photon energy, (b) output intensity, (c) spectral width, and (d) pulse width of DFB lasers with different grating periods ($\Lambda$) under a pumping power of 170 mW. Red squares and blue triangles denote simulation results for DFB lasers with as-cleaved facets and AR-coated facets, respectively, whereas yellow circles and lines represent experimental results. In (d), the pulse-width data at $\Lambda$ = 120 nm for the experiment and simulations with AR-coated and as-cleaved facets are 14.2, 9.7, and 9.1 ps, respectively, and are too large to be plotted.

Figure 7 shows the lasing photon energy (a), output intensity (b), spectral width (c), and pulse width (d) of the DFB lasers with different periods. Red squares and blue triangles denote simulation results, which are explained and discussed later in this paper. The yellow circles represent experimental results measured at a pumping power of 170 mW, which we discuss here.

Figure 7(a) shows the linear dependence of the photon energies of the output pulses from the DFB lasers on the grating period due to the Bragg condition. Figure 7(b) illustrates that the spectral peak intensity increases with the grating period. This is because the 124-nm-period DFB laser emits at a wavelength close to the optical gain peak of the quantum-well gain spectrum, whereas DFB lasers with shorter grating periods emit at shorter wavelengths, resulting in larger detuning from the gain peak and reduced amplification. Figure 7(c) shows that the corresponding spectral widths also increase with the grating period, except for the 120-nm-period DFB laser with dual-mode lasing. This indicates enhanced carrier-induced refractive-index modulation as the photon energy approaches the gain peak. Figure 7(d) exhibits that the 122-nm-period DFB laser produces the shortest output pulse width, whereas the pulse widths increase for grating periods both above and below 122 nm. Additionally, DFB lasers with larger grating periods exhibit higher output intensities and spectral widths, but they also have longer pulse widths than the 122-nm-period device.

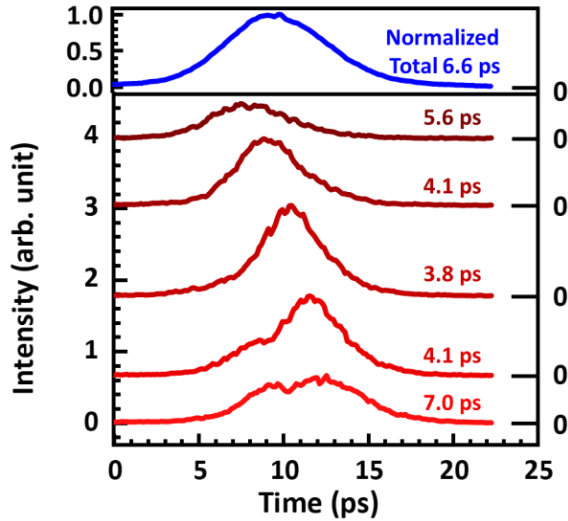

Fig. 8. Time traces at photon energies indicated by red arrows in Fig. 5(b) (data for $\Lambda$ = 122 nm at 170 mW) with an energy interval of 0.35 meV and normalized total-intensity temporal profile (top, blue).

Figure 8 shows spectrally resolved time traces at various photon energies, indicated by red arrows in Fig. 5(b), for the 122-nm-period DFB laser with a total pulse width of 6.6 ps. The time traces clearly indicate a down-chirp. Near the central photon energy, the pulse width reaches a minimum of 3.8 ps (2.3 ps after deconvolution), accompanied by the highest emission intensity. To the best of our knowledge, this is the shortest pulse width reported in gain-switched DFB lasers. In contrast, photon energies deviating from this central value show broader pulse widths and lower intensities. Notably, a side mode appears in the lowest-energy trace, further broadening the pulse width at lower photon energies.

The 122-nm-period DFB laser produced an output pulse with a width of 6.6 ps and a spectral width of 1.0 meV (corresponding to a frequency bandwidth of 256 GHz). The resulting time–bandwidth product was 1.7, exceeding the Fourier-transform-limited values of 0.44 for a Gaussian pulse and 0.31 for a sech$^2$ pulse, reflecting the influence of the down-chirp effect. The output pulse can be further shortened through spectral filtering near the central energy or by applying chirp compensation.

**SIMULATIONS**

We numerically calculated the time-dependent evolutions of photon density, carrier density, refractive index, and other dynamical variables in the DFB laser cavity during gain switching using the time-domain traveling-wave (TDTW) method [18–20,23]. In this approach, the coupled wave equations in Eq. (1) are solved for $F$ and $R$, which represent the amplitudes of the forward- and backward-propagating electric fields along the cavity ($z$-direction), respectively. The notation and parameter definitions used in these equations are provided in Appendix A.

$$\frac{1}{v_g}\frac{\partial F(z,t)}{\partial t} + \frac{\partial F(z,t)}{\partial z} = (\alpha - i\delta)F(z,t) + i\kappa R(z,t),$$

$$\frac{1}{v_g}\frac{\partial R(z,t)}{\partial t} - \frac{\partial R(z,t)}{\partial z} = (\alpha - i\delta)R(z,t) + i\kappa F(z,t). \quad (1)$$

For the local carrier density $N$, we adopted a phenomenological $ABC$ model of the nonlinear rate equation in Eq. (2). Here, $S$ is the local photon density, which is proportional to the square of the electric field amplitude in Eq. (1).

$$\frac{dN(z,t)}{dt} = \eta N_{pump}(z,t) - \Gamma \frac{c}{n_g}\frac{g}{m}S - AN - BN^2 - CN^3,$$

$$S(z,t) = \frac{\varepsilon_0 n_{eff}(z,t)}{2h\nu}\frac{c}{v_g}\left[|F(z,t)|^2 + |R(z,t)|^2\right]. \quad (2)$$

A nonlinear gain model [43] was also used (Eq. (3)), with the variation of the refractive index $n_{eff}$ modeled via linear approximation.

$$g(z,t) = A_0(N - N_{tr})\left[1 + \frac{A_0(N - N_{tr})}{g_s}\right]^{-1}(1 + \varepsilon S)^{-1},$$

$$\alpha(z,t) = \frac{\Gamma g - \alpha_{loss}}{2}, \quad (3)$$

$$n_{eff}(z,t) = n_0 + \Gamma\frac{\partial n}{\partial N}N.$$

A sufficiently fine numerical discretization was used, with temporal and spatial resolutions of ~1 fs and 83 nm, respectively. Temporal waveforms at individual wavelengths were calculated by adding a continuously tunable bandpass filter at the output facet of the DFB cavity. We assembled the resulting temporal waveforms at discrete wavelengths into a two-dimensional time–wavelength map.

Simulations were performed for two cases: cavity facets coated with antireflection (AR) coatings and as-cleaved facets with a reflectivity of 0.3. An effective photon lifetime was estimated in the TDTW calculation with photon injection under zero-gain conditions. For the AR-coated facet, the photon lifetime was 0.4 ps for FP lasers and 0.8 ps for DFB lasers. For the as-cleaved facets, the photon lifetime increased to 1.4 ps for FP lasers and 1.5 ps for DFB lasers.

The pumping level $P_{low}$ ($P_{high}$) was set to 1.30 (1.56) times the threshold level of the 124-nm-period DFB laser and 1.17 (1.40) times that of the 122-nm-period device for both the as-cleaved and AR-coated facets. $P_{high}$ = 1.2 $P_{low}$ corresponded to an optical pumping level of 170 mW, which was equal to 1.2 times the experimental value of 143 mW.

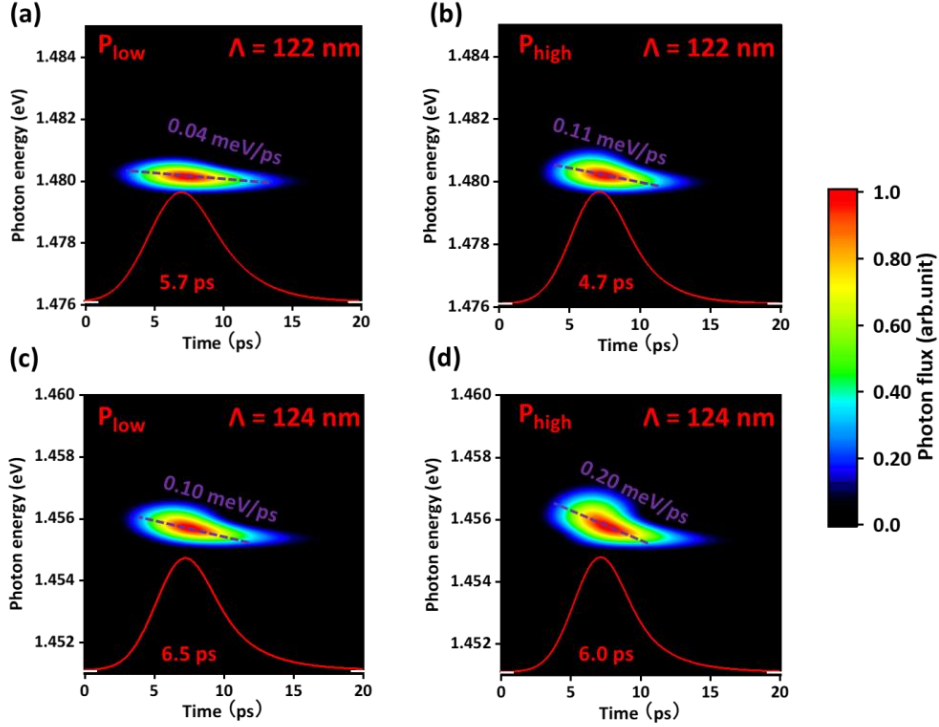

Fig. 9. Simulated time-resolved spectra of the DFB laser with as-cleaved facets and grating periods (a, b) $\Lambda$ = 122 nm and (c, d) $\Lambda$ = 124 nm. The pumping levels correspond to (a, c) 143 mW ($P_{low}$) and (b, d) 170 mW ($P_{high}$) or (a) 1.17, (b) 1.40, (c) 1.30, and (d) 1.56 times higher pumping than the threshold level $P_{th}$.

Figure 9 shows the simulated time-resolved spectra of SG-DFB lasers with as-cleaved facets having a reflectivity of 0.3. As shown in Fig. 9(a), at a lower pumping power of 143 mW, corresponding to the experimental results in Fig. 5(a), the 122-nm-period DFB laser generates a gain-switched pulse with a width of 5.7 ps and a down-chirp rate of 0.04 meV/ps. When the pumping power increases to 170 mW, the output pulse width shortens to 4.7 ps and the chirp rate increases to 0.11 meV/ps, as shown in Fig. 9(b).

Figures 9(c) and 9(d) show the simulation results for the 124-nm-period DFB laser at pump powers of 143 and 170 mW, respectively. The time-resolved spectra of the 124-nm-period DFB laser exhibited broader spectra than those of the 122-nm-period DFB laser. The gain-switched pulses had widths of 6.5 and 6.0 ps, both broader than those of the 122-nm-period DFB laser, despite its higher output power. The chirp rate of the 124-nm-period DFB laser also increased from 0.10 to 0.20 meV/ps with higher pump power, resulting in shorter pulse durations and broader spectral widths.

The observed down-chirp reflected the time-dependent refractive-index variation associated with carrier-density depletion. At the onset of the gain-switched pulse, the carrier density was high, lowering the refractive index and raising the photon energy. As the pulse evolved, carrier depletion from stimulated emission increased the refractive index, shifting the emission toward lower photon energies.

Figure 7 shows the simulation results for the lasing photon energy (a), output intensity (b), spectral width (c), and pulse width (d) of DFB lasers with different grating periods in the as-cleaved and AR-coated cases, plotted as red squares and blue triangles, respectively. The pumping level in the simulation corresponds to the experimental optical pumping level of 170 mW.

The lasing photon energy (Fig. 7(a)) exhibited a linear relationship with the grating period in both cases, consistent with the experimental results related to the Bragg condition. Similarly, the normalized output intensity (Fig. 7(b)) increased with larger grating periods, consistent with the gain spectrum.

Figure 7(c) shows that the simulated spectral width increases with the grating period, consistent with the experimental data. The increased width reflected larger variations in carrier density and, hence, refractive index during lasing. In addition, the 120-nm-period DFB laser exhibited single-mode lasing in the simulation, whereas multimode lasing was observed in the experiment. Other spectrum-broadening effects present only in experiments included fluctuations in the grating period due to the limited accuracy of EBL and dry etching, as well as nonuniform pumping within the cavity.

Regarding the output pulse widths (as shown in Fig. 7(d)), the simulation results showed similar trends to the experimental data, although the experimental result demonstrated a larger pulse width due to the 3.0-ps time resolution. The AR-coated case generated a shorter pulse than the as-cleaved case, possibly due to higher mirror loss and a shorter photon lifetime. Notably, the 122-nm-period DFB laser produced the shortest pulse among all DFB laser periods. The 124-nm-period DFB laser produced a longer pulse width than 121–123-nm-period DFB lasers, even though the 124-nm-period device exhibited higher output intensity and spectral width.

The good agreement between simulation and experiment supports the validity of the theoretical model, parameters, and computations. This indicates that our simulation can be used for predicting pulse characteristics for similar gain-switched DFB lasers under optical and electrical pumping and for understanding the mechanisms or physics of sophisticated gain-switched DFB lasers.

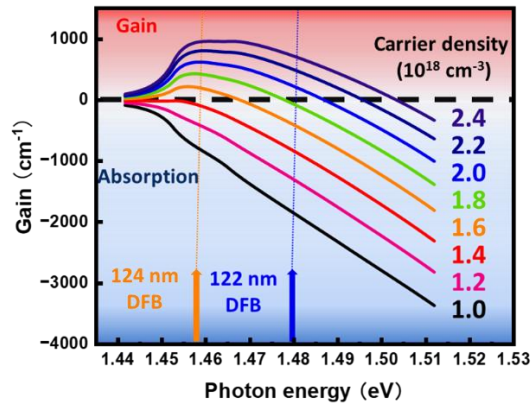

Fig. 10. Calculated gain spectra of the 10-period MQW structure at different carrier densities.

Figure 10 shows the calculated gain spectra under different carrier densities in the MQWs of $GaAs/Al_{0.33}Ga_{0.67}As$. The region above the black dashed line (red) represents gain, whereas the region below (blue) indicates absorption. The 122-nm-period DFB lasers (blue arrow) exhibit higher differential gain, saturation gain, and transparency carrier density in the higher-photon-energy region of the quantum-well gain spectrum than the 124-nm-period DFB lasers (orange arrow), notwithstanding the latter's highest peak gain in the gain spectrum. These properties increase the stimulated emission rate and accelerate carrier depletion, ultimately leading to faster gain switching and faster absorption recovery, resulting in shorter pulse generation.

Finally, this study establishes design principles and suitable conditions for short-pulse generation in gain-switched DFB lasers and computational simulation tools. With spectral filtering or chirp compensation, a pulse width of 3.0 ps or shorter becomes achievable, positioning gain-switched DFB lasers as promising sources for a wide range of applications.

**CONCLUSION**

We designed and fabricated optically pumped SG-DFB lasers with varying grating periods to investigate gain-switching dynamics. Gain-switched pulses were generated via 200-fs impulsive optical excitation at room temperature and measured using a spectrometer and a streak camera.

In time-resolved spectral measurements, a shorter pulse duration of 6.6 ps with a chirp rate of 0.13 meV/ps was achieved using a 122-nm-period DFB laser under 170 mW of optical pumping. In contrast, devices with grating periods other than 122 nm produced broader pulses, even though the 124-nm-period DFB laser produced higher output intensity due to higher gain. After spectral filtering, the minimum pulse width of 2.3 ps (deconvolved from 3.8 ps with a 3.0-ps time resolution, including inherent time jitter) at the central photon energy component was obtained.

We also developed a TDTW method to simulate carrier, photon, gain, and refractive-index dynamics within the DFB laser cavity and to directly compute time-resolved spectra for different grating periods in both AR-coated and as-cleaved (0.3 reflectivity) configurations. Time-resolved measurements and TDTW simulations revealed an excitation-power-dependent down-chirp effect in DFB lasers, originating from carrier-density-induced refractive-index variation during pulse generation.

Both measurements and simulations showed that the lasing wavelength, spectral width, and output intensity of DFB lasers increased as the grating period increased monotonically from 120 to 124 nm. Notably, the 122-nm-period DFB laser generated the shortest pulses given its higher saturation gain, differential gain, and transparency carrier density in the higher-photon-energy region of the quantum-well gain spectrum.

The proposed mechanisms provide practical design guidelines for achieving short-pulse generation in gain-switched DFB lasers. The TDTW simulation is also a useful tool for ultrafast semiconductor lasers and integrated photonic platforms. Altogether, they enable the future development of high-speed optical communications, ultrafast spectroscopy, and on-chip photonic systems. Furthermore, the nearly linear down-chirp in the single-mode gain-switched SG-DFB laser indicates that even shorter pulses of about 3.0 ps can be obtained through chirp compensation or spectral filtering techniques.

**APPENDIX A: SIMULATION PARAMETERS**

In Eq. (1), $\alpha$ is the net gain coefficient for the electric-field amplitude, $\kappa$ is the coupling coefficient, $v_g$ is the group velocity of light, and $\delta$ is the Bragg wavevector detuning.

In Eq. (2), $\eta$ is the pumping efficiency, $\Gamma = 0.31$ is the optical confinement factor, $m = 10$ is the number of quantum wells, $n_g = 3.697$ is the group refractive index, and $g$ denotes material gain. We used $A \approx 7 \times 10^7 \, s^{-1}$ as the nonradiative recombination coefficient, $B \approx 2.7 \times 10^{-16} \, m^3 s^{-1}$ as the radiative recombination coefficient, and $C \approx 2.9 \times 10^{-42} \, m^6 s^{-1}$ as the Auger recombination coefficient.

In Eq. (3), the parameters are defined as follows: saturation gain $g_s$, differential gain coefficient $A_0$, refractive index without carrier injection $n_0 = 3.44$, nonlinear gain coefficient $\varepsilon = 1.5 \times 10^{-23}$, internal loss $\alpha_{loss} = 4000 \, m^{-1}$, and transparency carrier density $N_{tr}$. Carrier injection is modeled as a Gaussian-shaped excitation pulse with a pulse width of 200 fs.

The parameters of our present SG-DFB lasers, including the coupling coefficient, optical confinement factor, gain spectrum, *ABC* coefficients, cavity loss, and so on, are directly calculated using the commercial software PICS3D (Crosslight Software Inc.), based on the device structure used in the experiment (Table A1).

**Table A1.**

| Parameters for different grating periods of DFB lasers | | | | | |
|---|---|---|---|---|---|
| $\Lambda$ (nm) | 120 | 121 | 122 | 123 | 124 |
| $\partial n/\partial N$ ($10^{-27}$ m$^3$) | −7.21 | −7.07 | −7.5 | −8.57 | −11.21 |
| $A_0$ ($10^{-19}$ m$^2$) | 1.4 | 1.52 | 1.64 | 1.5 | 1.17 |
| $g_s$ ($10^5$ m$^{-1}$) | 5.74 | 5.74 | 5.68 | 4.85 | 3.44 |
| $N_{tr}$ ($10^{24}$ m$^{-3}$) | 2.22 | 1.92 | 1.68 | 1.53 | 1.37 |

**Funding.** This work was partly supported by CREST (JPMJCR2544) from JST, FORWARD (# JPMI250310003) from MIC, JSPS-NSFC Joint Research Program (JPJSBP120227402), KAKENHI (JP23K13039 and JP24K00919) from JSPS, and Q-LEAP from MEXT, ARIM at NIMS from MEXT (JPMXP1224NM0008), Space Strategy Fund (SSF) from JAXA (JPJXSSF24MX17003), the AMADA Foundation, the Research Foundation for Opto-Science and Technology (REFOST), and UTokyo-JAXA collaboration project, in Japan.

**Acknowledgments.** We thank Drs. Eiichiro Watanabe, Hirotaka Oosato, Naoki Ikeda, and Michiko Fujii of NIMS for assistance in advanced nanofabrication.

**Disclosures.** The authors declare no conflicts of interest.

**Data availability.** Data underlying the results presented in this paper are not publicly available at this time but may be obtained from the authors upon reasonable request.